\documentclass[aoas,nameyear,dvips]{arximspdf}
\usepackage{graphics}
\doi{10.1214/09-AOAS292}
\volume{4}
\issue{1}
\pubyear{2010}
\firstpage{203}
\lastpage{221}

\makeatletter
\def\E{\mathrm{E}}

\def\pr{\operatorname{pr}}
\def\calN{\mathcal{N}}

\def\dto{\stackrel{d}{\longrightarrow}}
\def\Real{\mathbb{R}}

\newcommand{\eqref}[1]{(\ref{#1})}
\newtheorem{theorem}{Theorem}[section]
\makeatother

\begin{document}
\begin{frontmatter}

\title{Model misspecification in peaks over\\ threshold analysis}
\runtitle{Misspecification in peaks over thresholds}

\begin{aug}
\author{\fnms{M\'{a}ria} \snm{S\"{u}veges}\ead[label=e1]{Maria.Suveges@epfl.ch}}\and
\author{\fnms{Anthony C.} \snm{Davison}\thanksref{t1}\ead[label=e2]{Anthony.Davison@epfl.ch}\corref{}}
\pdfauthor{Maria Suveges, Anthony C. Davison}
\thankstext{t1}{Supported in part by the Swiss National
Science Foundation and the CCES project EXTREMES (\protect\url{http://www.cces.ethz.ch/projects/hazri/EXTREMES}).}
\affiliation{Ecole Polytechnique F\'{e}d\'{e}rale de Lausanne}
\runauthor{M. S\"{u}veges and A. C. Davison}
\address{Ecole Polytechnique F\'{e}d\'{e}rale\\
\quad de Lausanne\\
EPFL-FSB-IMA-STAT \\
Station 8, 1015 Lausanne\\
Switzerland\\
\printead{e1}\\
\phantom{E-mail: }\printead*{e2}}
\end{aug}

% HISTORY:
\received{\smonth{7} \syear{2009}}
\revised{\smonth{9} \syear{2009}}

% ABSTRACT
%
\begin{abstract}
Classical peaks over threshold analysis is widely used for statistical
modeling of sample extremes, and can be supplemented by a model for the
sizes of clusters of exceedances. Under mild conditions a compound
Poisson process model allows the estimation of the marginal
distribution of threshold exceedances and of the mean cluster size, but
requires the choice of a threshold and of a run parameter, $K$, that
determines how exceedances are declustered. We extend a class of
estimators of the reciprocal mean cluster size, known as the extremal
index, establish consistency and asymptotic normality, and use
the compound Poisson process to derive misspecification tests of model
validity and of the choice of run parameter and threshold. Simulated
examples and real data on temperatures and rainfall illustrate the
ideas, both for estimating the extremal index in nonstandard situations
and for assessing the validity of extremal models.
\end{abstract}

% KEYWORDS
%

\end{frontmatter}

%s1 ###
\section{Introduction}\label{sec1}

When extreme-value statistics are applied to time series it is common
to proceed as though the data are independent and identically
distributed, although they may be nonstationary with complex covariate
effects and with rare events generated by several different mechanisms.
Moreover, models that are mathematically justified only as asymptotic
approximations may be fitted to data for which these approximations are
poor. In this paper we suggest diagnostics for failure of these models
and illustrate their application.

A standard approach to modeling the upper tail of a distribution is the
so-called peaks over threshold procedure [\citet{davisonsmith}], under
which a threshold $u$ is applied to data $x_1,\ldots, x_n$ from a
supposedly stationary time series, leaving~$N$ positive exceedances
$x_j-u$. Extrapolation beyond the tail of the data is based on a fit of
the generalized Pareto distribution [\citet{pickands75}]
%
%e1 ###
\begin{equation}\label{gpd}
H(y) =
\cases{
1-(1+\xi y/\sigma)^{-1/\xi}, &\quad$\xi\neq0$,\vspace*{2pt}\cr
1 - \exp(-y/\sigma),&\quad$\xi=0$,
}
\end{equation}
to the $N$ exceedances, treated as independent.
The parameters in (\ref{gpd}) are a scale parameter $\sigma>0$ and a shape
parameter $\xi\in\Real$ that controls the weight of the tail of the distribution, whose
$r$th moment exists only if $r\xi<1$. In many applications $\xi$
appears to lie in the interval $(-0.5, 0.5)$, but uncertainty about its
value generally leads to alarmingly wide confidence intervals for
quantities of interest such as a return period $T$ or a return level
$x_T$; these satisfy $\pr(X>x_T) = 1/T$. If the time series is white
noise, $n$ is large and $N/n$ is small, then exceedances of $u$ appear
as a Poisson process, and under mild conditions we may use the tail
approximation $\pr(X>x) \approx(N/n) \{1-\hat H(x-u)\}$, where $\hat
H$ is the estimate of (\ref{gpd}). A crucial preliminary to using such
methods is the choice of threshold $u$, which is usually performed
graphically using stability properties of (\ref{gpd}): if $Y\sim H$ and
$u>0$, then conditional on $ Y>u$ the exceedance $Y-u$ has distribution~(\ref{gpd}) with parameters~$\xi$ and $\sigma' = \sigma+ \xi u$; and
if $\xi<1$, then the mean residual life $\E(Y-u\mid Y>u)=\sigma'/(1-\xi
)$. It is standard practice to plot empirical versions of these
quantities for a range of potential thresholds, and to use only values
of $u$ above which the estimates appear to be stable and the empirical
mean residual life appears to be linear. See Coles [(\citeyear{colesbook}), Chapter~4] or
\citet{statofext} for more details, and
\citet{BeirVyncTeugexce1996} and \citet{SousMichdiag2004} for variants
of the last plot intended to stabilize it in the presence of
heavy-tailed data.

This approach to tail modeling is based on a general and well-developed
probabilistic theory of extremes [\citet{leadbetter},
\citet{embrechtsetal}, \citet{falketalbook}] and is widely used: in 2008
alone the Web of Science records around 450 articles in which the terms
``peaks over threshold'' or ``generalised Pareto distribution'' appear in
the abstract or title. Thus, it is important to develop simple tools
for diagnosis of the failure of such models.

One source of failure is the choice of threshold. A bad choice may
yield a poor tail approximation, both because the generalized Pareto
distribution is inappropriate if $u$ is too small and because
independence assumptions used to fit the model are invalid: in
practice, the observations, and therefore the exceedances, are almost
always dependent. This dependence is often reduced by declustering the
exceedances, for example, declaring that those lying closer together
than a run parameter $K$ belong to the same cluster, and fitting (\ref{gpd}) only to the largest exceedance of each cluster. However, a poor
choice of $K$ will give a poor inference, so it is essential to check
how the results depend on the choices of threshold $u$ and run
parameter $K$.

A key issue is thus the effect of possible model misspecification on
inference. \citet{whitebook} gives conditions under which the maximum
likelihood estimator derived from a misspecified model is a consistent
and normally distributed estimator of the parameter that minimizes the
Kullback--Leibler discrepancy between the true and the assumed models,
and constructs tests for misspecification. In the context of statistics
of extremes, the peaks over the threshold model may be justified by a
compound Poisson process model for the exceedances of a random process
above a high threshold [\citet{hsing87}, \citet{hsingetal88}]. This model, which
we outline in Section~\ref{subseclik}, can be checked through the projection of the
two-dimensional limiting point process of exceedances onto the time
axis: if the projection is misspecified, we should be wary about using
peaks over thresholds.

The main contribution of this paper is to construct diagnostics for the
adequacy of peaks over threshold models.
Section~\ref{sectheor} introduces inter-exceedance times truncated by
the run parameter $K$, which we call $K$-gaps, and discusses the
selection of an appropriate run parameter and threshold. We propose the
use of an information sandwich as a diagnostic for model failure, and,
as a byproduct, we extend the maximum likelihood estimator of the
reciprocal mean cluster size, the extremal index, given in \citet{suveges07}. Section~\ref{secsimu} uses data simulated from two
autoregressive models and a Markov chain model to illustrate the
application of our ideas. Section~\ref{secdata} applies them to real
data, to elucidate nonstationarity and tuning parameter selection and
to aid extremal index estimation when the basic assumptions of
extreme-value theory are violated. Section~\ref{secdisc} contains a
brief discussion.

%s2 ###
\section{Theory}\label{sectheor}

%s2.1 ###
\subsection{Likelihood}\label{subseclik}

We consider asymptotic models for the upper extremes of a strictly
stationary random sequence $X_1, \ldots, X_n$ with marginal
distribution function $F$. A standard approach is to consider the
limiting point process of rescaled variables $\calN_n = \sum_{i=1}^n
\delta_{i/n, (X_j-b_n)/a_n}$ as $n\to\infty$, where the sequences $\{
b_n\}\subset\Real$ and $\{a_n\}\subset\Real_+$ are chosen so that the
maximum $a_n^{-1}(\max\{ X_i\}-b_n)$ has a nondegenerate limiting
distribution $G$ [\citet{resnick87}]. Under mild conditions, if $\calN_n
$ converges in distribution to a point process $\calN$ as $n\to\infty$,
this must have the representation [\citet{hsing87}]
\[
\calN= \sum_{i = 1}^{\infty} \sum_{j = 1}^{M_i} \delta_{(S_i, X_{ij})},
\]
where $(S_i, X_{i1})$ are the points of a nonhomogeneous Poisson point
process with mean measure $|\cdot| \times\tau(\cdot)$ on $[0, 1)
\times(x_L,x_R]$, $|\cdot|$ is the Lebesgue measure, $x_L$ and $x_R$
are the left and right endpoints of $G$ and $\tau(x, \infty] = - \log
G(x)$. The $X_{ij}$ are such that, for all $i$, the variables
\[
Y_{ij} = {-\log G(X_{ij})\over-\log G(X_{i1})}, \qquad j = 1, \ldots, M_i,
\]
are the points of a point process $\gamma_i$ on $[1,\infty)$ with an
atom at unity. The $\gamma_i$ are independent of the nonhomogeneous
Poisson process $(S_i, X_{i1})$ and of each other, and are identically
distributed. Thus, the point process limit of $\calN_n$ is a compound
Poisson process $\calN$ comprising independent identically distributed
clusters, the $i$th of which has $M_i$ exceedances $Y_{i1}, \ldots,
Y_{iM_i}$ that may have different sizes but occur simultaneously.

This result implies convergence in distribution of maxima to the
generalized extreme-value distribution, convergence of threshold
exceedances to the generalized Pareto distribution, and convergence of
the projection $\calN^*_n = \sum_{i=1}^n \delta_{i/n}$ to a compound
Poisson process with points at $S_i$ and marks $M_i$. In the limit the
inter-exceedance times follow a mixture of an exponential distribution
and a point mass at zero [\citet{ferro03}], and this remains true for
inter-exceedance times truncated by some fixed positive value; see
below. For a sequence of thresholds~$u_n$, define the inter-exceedance
times in the sequence $\{X_i\}$ by
\[
T(u_{n}) = \min\{k \geq 1{}\dvtx{} X_{k+1} > u_n   |   X_1 > u_n\},
\]
and the corresponding $K$-gaps by
\[
S^{(K)}(u_{n}) = \max \{T(u_{n})-K,0 \},\qquad  K = 0, 1,
\ldots.
\]
Then Theorem 1 of \citet{ferro03} can be modified to yield a limiting
distribution for the $K$-gaps, which \citet{suveges07} gave for $K=1$.
The proof requires only small modifications of the original, by
considering $\pr\{\overline{F}(u_n)[T(u_n) - K] > t\}$, where $\overline
{F}(u_n) = 1 - F(u_n)$.
Let $\mathcal{F}_{i,j}(u_{n})$ denote the $\sigma$-field generated by
the events $X_r \leq u_n, r = i, \ldots, j$. For any $\mathcal{A} \in
\mathcal{F}_{1,k}(u_{n})$ with $\pr(\mathcal{A})> 0$, $\mathcal{B} \in
\mathcal{F}_{k+l,n}(u_{n})$ and $k,l$ integers such that $k = 1, \ldots
, n-l$, define
\[
\alpha^{*} (n,l) = \max_{k} \sup_{\mathcal{A}, \mathcal{B}} |\pr
(\mathcal{B} \mid \mathcal{A}) - \pr(\mathcal{B})|.
\]
Then we have the following result.
%t2.1
\begin{theorem} \label{theo_Kgapconv}
Suppose there exist sequences of integers $ \{ r_{n} \}$
and of thresholds $\{u_{n}\}$ such that as $n \rightarrow\infty$, we
have $r_{n} \rightarrow\infty$,
$r_{n}\overline{F}(u_{n}) \rightarrow\tau$
and $\pr \{M_{r_{n}} \leq u_{n}  \} \rightarrow
e^{-\theta\tau}$ for some $\tau\in
(0, \infty)$ and $\theta\in(0,1]$. Moreover, assume that there
exists a sequence $l_{n} = o(n)$ for
which $\alpha^{*} (cr_{n},l_{n}) \rightarrow0$ as $n \rightarrow
\infty$ for all $c > 0$.
Then as $n \rightarrow\infty$,
%
%e2 ###
\begin{eqnarray} \label{mxtexp}
\pr \bigl\{\overline{F}(u_{n})S^{(K)}(u_{n})
> t \bigr\} \longrightarrow\theta\exp(-\theta t),\qquad  t>0,
\end{eqnarray}
where the extremal index $\theta$ lies in the interval $(0,1]$ and is
the reciprocal of the mean cluster size, that is, $\E(M_i)=\theta^{-1}$.
\end{theorem}

Equation~(\ref{mxtexp}) corresponds to a limiting mixture model for the
intervals between successive exceedances: with probability $\theta$ the
interval is an exponential variable with rate $\theta$, and otherwise
it is of length zero, yielding a compound Poisson process of exceedance
times and a likelihood-based estimator of $\theta$. Suppose that $N$
observations from a stationary random sequence $X_{1}, \ldots, X_n$
exceed the threshold $u_{n}$, let the indices $\{j_i {}\dvtx{}X_{j_i}> u_n\}$
denote the locations of the exceedances, let $T_i = j_{i+1} - j_{i}$
denote the inter-exceedance times, and let $S_i^{(K)} = \max(T_i -
K,  0)$ denote the $i$th $K$-gap, for $ i = 1, \ldots, N-1$ and $K =
0, 1, \ldots.$
Assuming independence of the gaps $S_1^{(K)},\ldots, S_{N-1}^{(K)}$,
the limiting distribution~(\ref{mxtexp}) leads to the log likelihood
%
%e3 ###
\begin{eqnarray} \label{goodloglik}
\hspace*{25pt}\ell_K \bigl(\theta; S_i^{(K)}  \bigr) =
(N-1-N_C) \log(1-\theta) + 2 N_C \log\theta- \theta\sum
_{i=1}^{N-1} \overline{F}(u_n) S_{i}^{(K)}
\end{eqnarray}
for $\theta$, where $N_C = \sum_{i=1}^{N-1} I(S_{i}^{(K)} \neq0)$, and to a closed-form\vspace*{-1pt}
maximum likelihood estimator $\hat{\theta}_n$, which
is the smaller root of a quadratic equation. Below we modify the log
likelihood~(\ref{goodloglik}) to allow nonstationarity to be detected
by using smoothing [\citet{fangijbels}, \citet{suveges07}].

%s2.2 ###
\subsection{Model misspecification}

The point process approach can fail because the assumptions of strict
stationarity and independence at extreme levels are violated, but even
if they are fulfilled, the chosen threshold parameter $u_n$ and the run
parameter $K$ may be inappropriately small, thereby leading to a poor
extreme-value approximation or to dependent threshold exceedances. In
order to detect such difficulties, we turn to classical work on model
misspecification [\citet{white82}]. Under broad assumptions, the maximum
likelihood estimator derived from a misspecified likelihood $\ell(\theta
)$ exists as a local maximum of $\ell(\theta)$. When the true model
$\ell_0$ is not contained in the postulated model family, that is,
there is no $\theta_0$ such that $\ell_0 = \ell(\theta_0)$, this
estimator is consistent for that parameter value $\theta_*$ within the
misspecified family $\ell(\theta)$ that minimizes the Kullback--Leibler
discrepancy with the true distribution. Define
$J(\theta) = \E_0 \{ \ell'(\theta, S_j)^2 \}$, $I(\theta) = - \E_0\{
\ell''(\theta, S_j) \}$, where the prime denotes differentiation with
respect to $\theta$, and $\E_0$ means expectation under the true model,
and their empirical counterparts
\[
\bar{J}_n(\theta) = (N-1)^{-1} \sum_{j = 1}^{N-1} \ell'(\theta, S_j)^2,\qquad
\bar{I}_n(\theta) = - (N-1)^{-1} \sum_{j = 1}^{N-1} \ell''(\theta, S_j).
\]
{\spaceskip=0.166em plus 0.05em minus 0.02em
Under regularity conditions satisfied by the limiting distribution~\eqref{mxtexp} when~$0<\break\theta<1$,
}
Theorem 3.2 of \citet{white82} yields that, as $n\to\infty$,
\[
\sqrt{n}(\hat{\theta}_n - \theta_*) \dto{N} \{0, I(\theta_*)^{-2}J(\theta_*)  \},\qquad
\bar{I}_n(\hat{\theta}_n)^{-2} \bar{J}_n(\hat{\theta}_n) \stackrel{\mathrm{a.s.}}{\longrightarrow} I(\theta_*)^{-2} J(\theta_*),
\]
where $\dto$ and $\stackrel{\mathrm{a.s.}}{\longrightarrow}$ denote
weak and almost sure convergence, respectively. Thus, the estimator
derived from \eqref{goodloglik} using an arbitrary run parameter $K$ is
consistent for the value $\theta_*$ minimizing the Kullback--Leibler
divergence with the true distribution, and is asymptotically normally
distributed with the sandwich variance $I(\theta_*)^{-1} J(\theta_*)
I(\theta_*)^{-1}$, which can be estimated by its empirical counterpart
evaluated at $\hat{\theta}_n$.

It is straightforward to verify that the above theory applies to the
log likelihood~(\ref{goodloglik}), so that as $u_n$ increases in such a
way that $N\to\infty$, the corresponding maximum likelihood estimator
is consistent and asymptotically normal.

%s2.3 ###
\subsection{Diagnostics: the information matrix test} \label{subsecdiag}

Tests to detect model misspecification may be based on the fact that
the Fisher information for a well-specified regular model equals the
variance of the score statistic, that is, $J(\theta) = I(\theta)$. If
we write [\citet{white82}]
%
%e4 ###
\begin{eqnarray} \label{misspec_teststat}
D(\theta) &=& J(\theta) - I(\theta),
\end{eqnarray}
then one possible misspecification test amounts to testing the null
hypothesis $H_0: D(\theta) = 0$ against the alternative $H_1
{}\dvtx{}   D(\theta) \neq0$. Let $d(s_j,\theta)$ denote the one-observation
version of $D(\theta)$, let $D_{n}(\theta) = n^{-1} \sum_{j = 1}^n
d(s_j,\theta)$ denote the empirical counterpart of $D(\theta)$, and let
\[
V(\theta) = \E \{ [ d(S_j, \theta) + D{'}(\theta) I(\theta)^{-1}
\ell'(\theta, S_j) ]^2  \}
\]
and $V_n(\theta)$ denote the asymptotic variance of $D(\theta)$ and its
empirical counterpart. Detailed formulae are given in the \hyperref[appendix]{Appendix}.
Under mild regularity conditions, \citet{white82} proves the following
theorem, here given for a scalar parameter.
%t2.2
\begin{theorem} \label{theo_infotest}
If the assumed model $\ell(S_j, \theta)$ contains the true model for
some $\theta= \theta_0$, then as $n \rightarrow\infty$, $\sqrt{n}
D_{n}(\hat{\theta}_n) \dto{N} (0, V(\theta_0) )$,\vspace*{1.5pt}
$V_{n}(\hat{\theta}_n) \stackrel{a.s.}{\longrightarrow} V(\theta
_0)$, and
the test statistic $T(\hat{\theta}_n) = nD_{n}(\hat{\theta}_n)^2
V_{n}(\hat{\theta}_n)^{-1}$ is distributed as $\chi^2_1$.
\end{theorem}

To check the quality of this chi-squared approximation, we performed
simulations from the $\operatorname{AR}(1)$ and $\operatorname{AR}(2)$
processes described in Section~\ref{secsimu}, using choices of threshold and run parameter under which the
models are well specified. Probability plots of the simulated $T(\hat
{\theta}_n)$ showed that the $\chi^2_1$ approximation is good for
$N\geq80$, and tends to be conservative if $N<80$. Thus, relying on
this approximation can lead to a loss of power when the number of
exceedances is small, in which case it is difficult to detect
misspecification anyway. Below we shall use the chi-squared quantiles
without further comment.

%s3 ###
\section{Simulated examples} \label{secsimu}

%f1 ###
\begin{figure}

\includegraphics{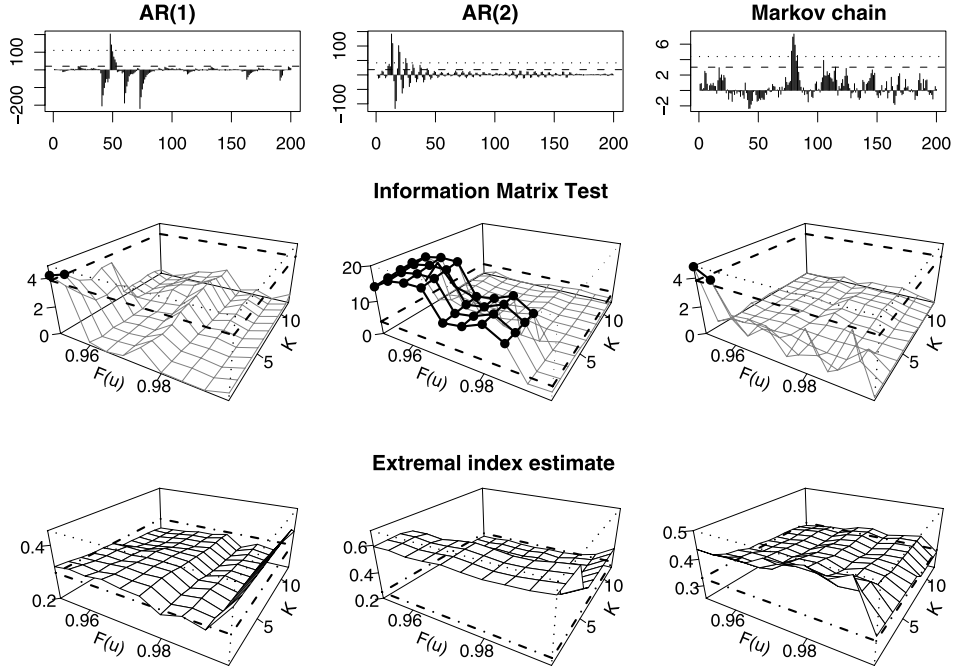}

\caption{Illustration of the
diagnostics, based on data simulated from an $\operatorname{AR}(1)$ model (left column),
an $\operatorname{AR}(2)$ model (middle column) and a Markov chain (right column).
The top row gives an impression of each series, together with its 0.95
and 0.99 quantiles (dashed and dotted horizontal lines, respectively).
The second row shows the information matrix test $T(\hat{\theta})$
(gray surface) and its\vspace*{1pt} 5\% critical value $\chi^2_1(0.95) = 3.84$
(thick dashed black line around the box), with the values above\vspace*{-1pt} $3.84$
accentuated by black blobs. The third row shows the estimated extremal
index $\hat{\theta}$ (black surface) as a function of the run parameter
$K$ and threshold $u$, with the true value of $\theta$ (thick
dash-dotted black line).}\label{figexamples}
\end{figure}

For a numerical assessment of the ideas in Section~\ref{sectheor}, we apply them to
three processes:
\begin{longlist}
\item[$\operatorname{AR}(1)$:] $Y_i = \phi Y_{i-1} + Z_i$ with $\phi= 0.7$ and $Z_i$
standard Cauchy, with $K = 1$ and $\theta= 0.3$;
\item[$\operatorname{AR}(2)$:] $Y_i = \phi_1 Y_{i-1} + \phi_2 Y_{i-2} + Z_i$, with $\phi
_1 = 0.95$, $\phi_2 = -0.89$ and $Z_i$ Pareto with tail index 2, with
$K = 5$ and $\theta= 0.25$;
\item[Markov chain:] With Gumbel margins, a symmetric logistic
bivariate distribution for consecutive variables and dependence
parameter $r = 2$ [\citet{smith92}], with $\theta= 0.33$ and $K$ unknown.
\end{longlist}
For each process we generated series of length $n = 8000$ and obtained
sequences of inter-exceedance times;
%using thresholds at the 0.95 and the 0.98 quantiles
the top row of Figure~\ref{figexamples} shows a
short sample with a typical extreme cluster from each process. We then
calculated the maximum likelihood estimates $\hat{\theta}$ for
$K=1,\ldots, 12$ and thresholds corresponding to the $0.95, 0.955,
\ldots, 0.995$ quantiles. The second row of Figure~\ref{figexamples}
shows the resulting surfaces for the information matrix statistic
$T(\hat{\theta})$. The lowest panels show the estimated extremal index.
For the $\operatorname{AR}(1)$ process, the information matrix test suggests
misspecification for the combination of low thresholds with small run
parameter $K$, but this disappears when $u$ or $K$ is increased. For
the $\operatorname{AR}(2)$ process, the information matrix test indicates clear
misspecification for most thresholds when $K\leq5$, and there is then
also substantial overestimation of the extremal index. The information
matrix test for the Markov chain suggests that although
well-specifiedness cannot be rejected for $K = 1$ and 2, inference with
a larger run parameter will be more reliable. Correspondingly, the
extremal index estimate is closer to the true value for larger run parameters.

%f2 ###
\begin{figure}

\includegraphics{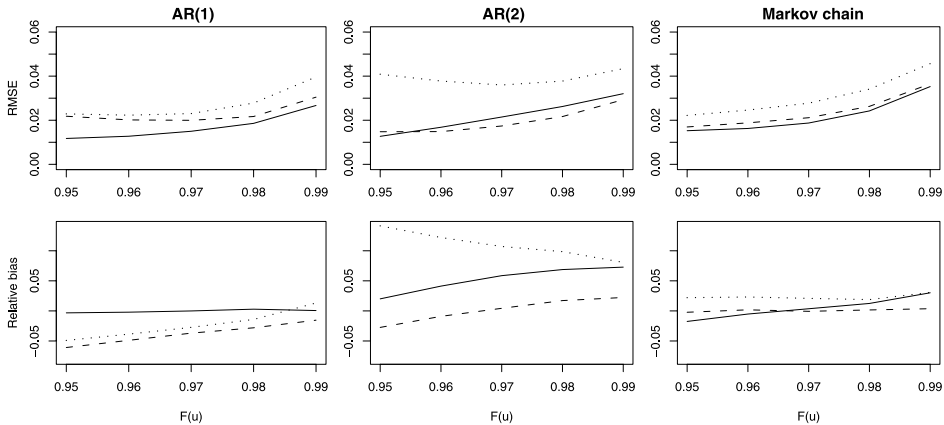}

\caption{Root mean squared error
(top row) and relative bias (bottom row) of the $K$-gaps maximum likelihood (solid),
the iterative least squares [S{\"{u}}veges (\protect\citeyear{suveges07}), dashed] and
the intervals (dotted) estimators on the $\operatorname{AR}(1)$ process (left panels),
on the $\operatorname{AR}(2)$ process (middle panels) and on the Markov chain (right
panels). $K = 1$ for the $\operatorname{AR}(1)$ process, $K = 6$ for the $\operatorname{AR}(2)$ process,
and $K = 5$ for the Markov chain. The number of observations was $n = 30{,}000$.}\label{figsimKgaps}
\end{figure}

To assess the quality of the extremal index estimator based on \eqref
{goodloglik}, a simulation study was performed with 1000 replications of
each of these processes, using \mbox{$K = 1$} for the $\operatorname{AR}(1)$ process, $K = 6$
for the $\operatorname{AR}(2)$, and $K = 5$ for the Markov chain as suggested by the
misspecification tests. We simulated processes of lengths $n = 2000$
and $n = 30{,}000$, and used thresholds corresponding to the 0.95, 0.96,
0.97, 0.98 and 0.99 quantiles. The median relative bias and the root
mean squared error for the case $n = 30{,}000$ are shown in Figure~\ref{figsimKgaps}. The plots confirm that if a suitable run parameter is
chosen, then the maximum likelihood estimator has generally lower bias
and root mean squared error than the most commonly used competitor, the
intervals estimator, and another good estimator based on an iterative
weighted least squares fit to the longest inter-exceedance times [\citet{suveges07}].

%f3 ###
\begin{figure}

\includegraphics{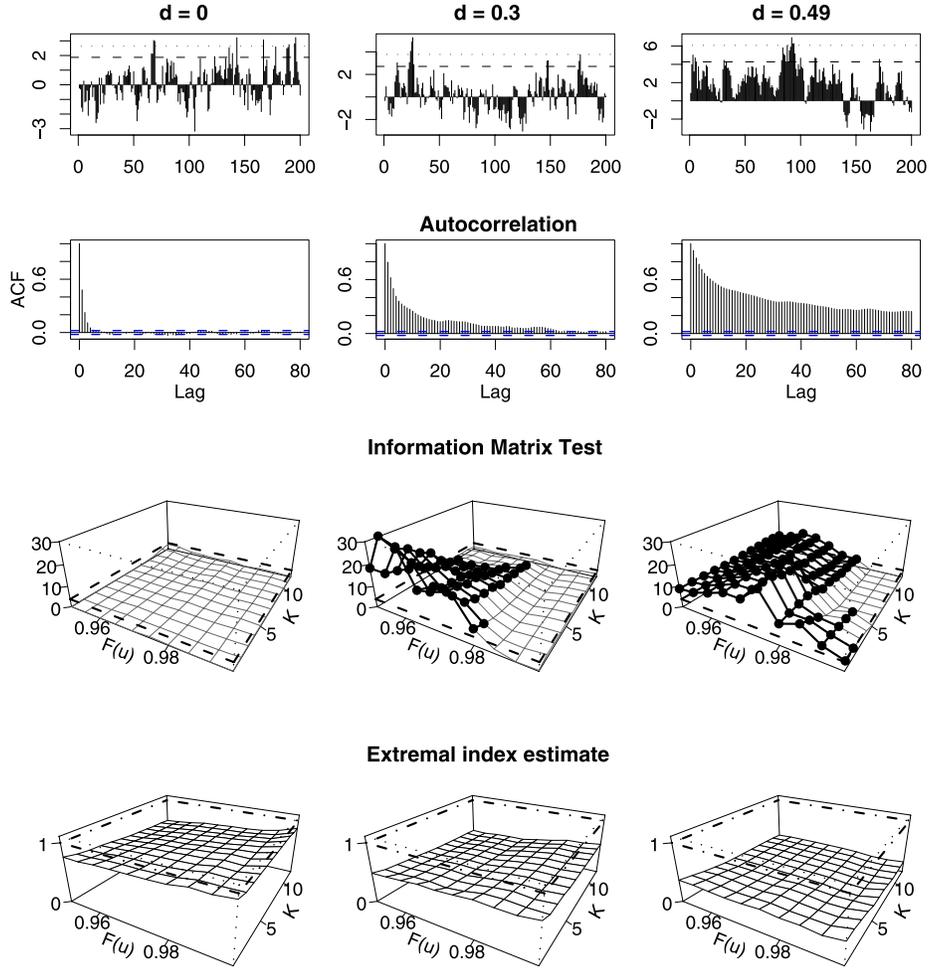}

\caption{The behavior of the
information matrix test as a function of the dependence range of the
process. The top row shows three $\operatorname{ARIMA}(1,0,d)$ sequences, with $d=0$
(left column), $d=0.3$ (middle column) and $d=0.49$ (right column), and
the second row their autocorrelation functions. The third row contains
the information matrix test $T(\hat{\theta})$ (gray surface),\vspace*{1pt} with the
thick dashed black lines indicating the critical value for the
information matrix test $\chi^2_1(0.95) = 3.84$,\vspace*{-1pt} and the black blobs
indicating $T(\hat{\theta}) > 3.84$. The fourth row presents the
extremal index estimate, with the thick dash-dotted lines representing
the theoretical value $\theta= 1$ for the case $d=0$.}\label{figlongrange}
\end{figure}

In order to explore the behavior of the misspecification tests in the
case of long-range dependence, we simulated fractionally differenced
$\operatorname{ARIMA}(1,0,d)$ processes of length $n=8000$, with Gaussian white noise
innovations, autoregressive parameter 0.5 and difference parameter
$d=0,0.3$ and $0.49$. The sequence with $d=0$ corresponds to a
stationary normal $\operatorname{AR}(1)$ process, and therefore has extremal index
$\theta= 1$. For the other cases, no theoretical calculations are
known to us concerning the existence of the extremal index. The
autocorrelation functions of the data, shown in the second row of
Figure~\ref{figlongrange}, show long memory when $d>0$. The test
statistic, plotted in the third row of Figure~\ref{figlongrange}, shows
a lengthening dependence range: misspecification is indicated at
thresholds up to $u=F^{-1}(0.985)$ and $K<8$ for $d=0.3$, and at all
thresholds and $K<8$ for $d=0.49$. The absence of misspecification at
very high thresholds for $d=0.3$ may be due to the effect of increasing
the threshold while keeping the length of the series fixed.

%s4 ###
\section{Data examples} \label{secdata}

%s4.1 ###
\subsection{Neuch\^{a}tel daily minimum summer temperatures} \label{subsecneu}

%f4 ###
\begin{figure}[b]

\includegraphics{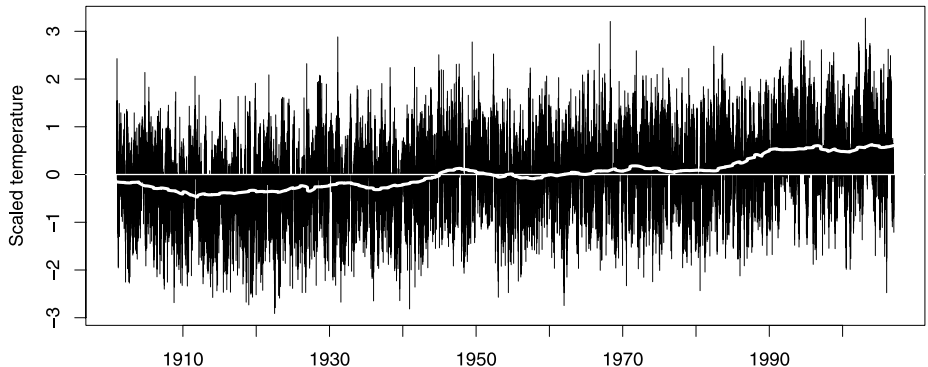}

\caption{The deseasonalized Neuch\^{a}tel daily minimum temperatures (black) with their trend, estimated
by a cubic spline-smoothed 10-year moving median (heavy white line).
The horizontal line is intended to help appreciate the trend.}\label{Tmintrend}
\end{figure}

For a first application to real data we use daily minimum summer
temperatures from Neuch\^{a}tel from 1 January 1901 to 31 May 2006.
Climatic change raises the question whether changes in temperature
extremes can be summarized simply by a smooth variation of the mean and
the variance of the entire temperature distribution, or whether there
are additional changes in the extremes. The daily summer minimum
temperatures at Neuch\^{a}tel show a strong trend in the averages, and
we investigate whether this is accompanied by a change in the
clustering of the extremes. The data have been carefully homogenized,
so such changes should not be due to changes in instrument siting or
type, or urban effects.

%f5 ###
\begin{figure}

\includegraphics{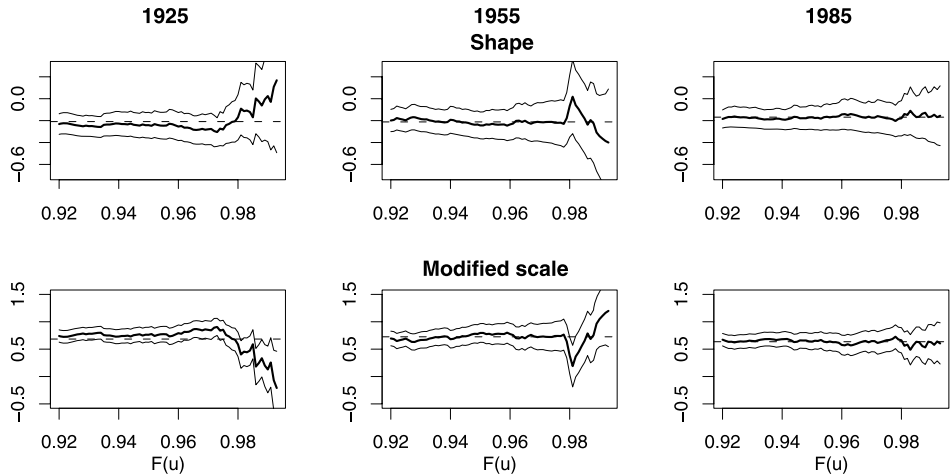}

\caption{Classical threshold
selection plots for three 41-year windows centered on 1925, 1955 and
1985 for the Neuch\^{a}tel daily minimum temperature anomalies, showing
the parameter estimate (bold) and pointwise 95\% confidence limits
(solid) as functions of the threshold. The dashed lines show the
average estimates for the different thresholds.}\label{classic3year}
\end{figure}

We stationarized the raw data by first centering and scaling by the
annual median and median absolute deviation~(MAD) cycle, and then
de-trending by using a ten-year moving median and MAD. Below we treat
the resulting standardized temperature anomalies for the months
June--August in successive years as a continuous time series. Figure
\ref{Tmintrend}, which shows the deseasonalized series before
de-trending, shows a strong irregular variation in the mean. The
presence of trend also motivates a careful misspecification analysis,
not only for appropriate estimation of the extremal index, but for the
assumption of stationarity.

Initially assuming stationarity of the anomalies in the 1901--2006
period, we applied the threshold selection procedures mentioned in
Section~\ref{sec1} and described in more detail by Coles [(\citeyear{colesbook}), Section~4.3] to the entire sequence. There seems to be stability
above the 0.98 quantile, and generalized Pareto analysis of the
complete sequence showed acceptable diagnostics. However, Figure~\ref{classic3year}, which shows these plots when applied separately to
three 41-year-long periods centered on the years 1925, 1955 and 1985,
casts some doubt on the overall results.

We therefore checked model misspecification as a function of time, by
centering 41-year long windows successively at 15 July of each year,
and calculating the information matrix test defined by equation \eqref
{misspec_teststat}, for every combination of threshold~$u$ and run
parameter $K$. The calculations thus gave 106 sets of $T(\hat{\theta})$
values, with extremal index estimates for all $(u,K)$ pairs for the
sequence of anomalies.

%f6 ###
\begin{figure}

\includegraphics{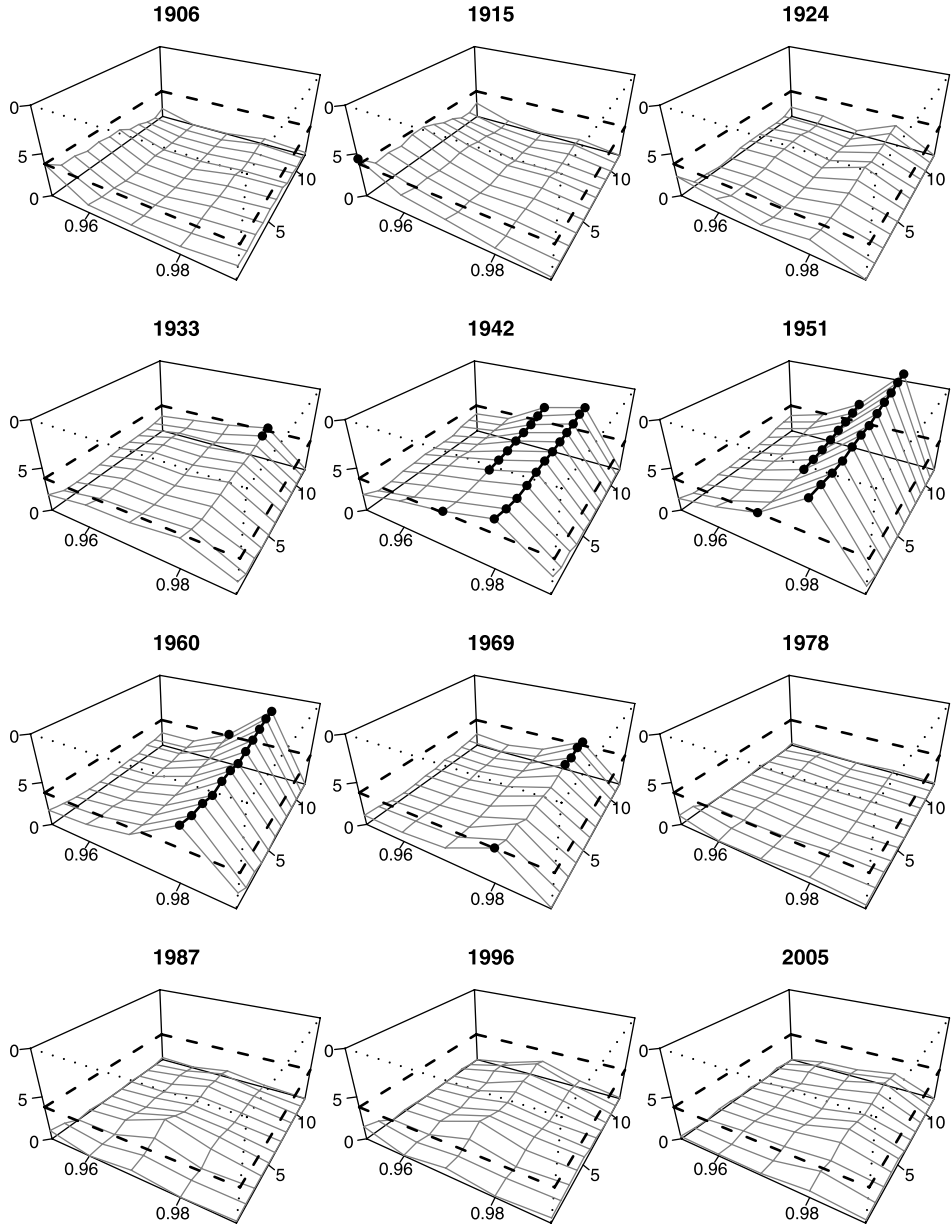}

\caption{Misspecification as a
function of time for the Neuchatel summer daily minimum temperature
anomalies. Horizontal foreground axis: threshold $u$ as $F(u)$;
horizontal left axis: run parameter $K$; vertical axis: $T(\hat{\theta})$.
The thick dashed lines around the box correspond to the critical
0.95-quantile of the $\chi_1^2$ distribution, the black blobs emphasize
the parameter combinations\vspace*{-1pt} where $T(\hat{\theta}) \geq\chi_1^2(0.95)$.
Years are indicated above the plots.}\label{pihistory}
\end{figure}

The three main potential sources of misspecification here are threshold
selection, the choice of run parameter and possible nonstationarity, so
the test statistics proposed above must depend on these. Figure~\ref{pihistory},\vspace*{1.5pt} which presents the surfaces of $T(\hat{\theta})$ for
twelve different years, suggests misspecification in the period
1935--1970 for thresholds around the 0.97- and 0.98-quantile for all
run parameters. Smaller instabilities were also found, mostly between
1985--2000, though these rarely exceeded the critical $\chi
^2_1$-quantile. These two periods roughly coincide with the strongest
nonstationarity in the summer mean temperatures: see Figure~\ref{Tmintrend}, in which the most marked periods of change in the 10-year
median are in the 1940s and in the 1980s. The threshold suggested by
classical selection methods, the 0.98-quantile, should be avoided: the
ridge indicating misspecification is located at this threshold.

%f7 ###
\begin{figure}

\includegraphics{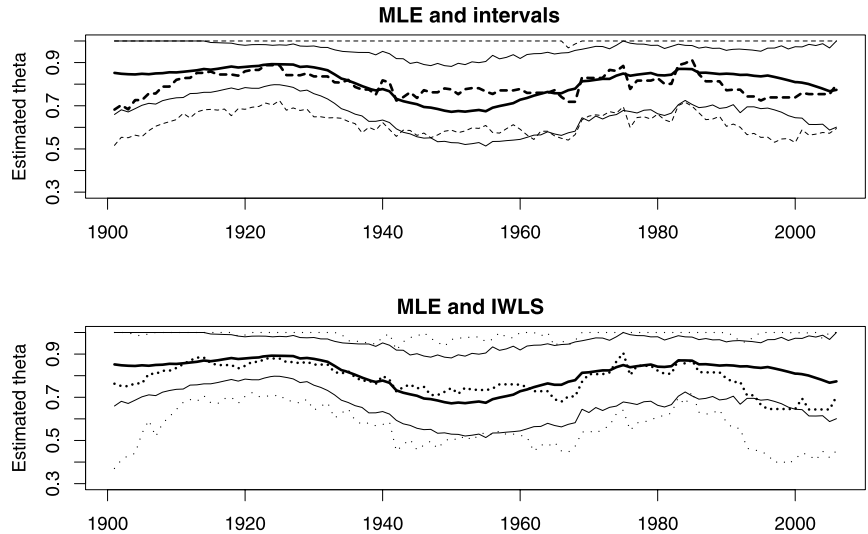}

\caption{Comparison of the maximum
likelihood estimator using $K = 4$ and $F(u) = 0.99$ (heavy solid in
both panels) to the intervals (top panel, thick dashed) and to the
iterative weighted least squares (bottom panel, thick dotted)
estimators. The 95\% confidence intervals are shown by thin versions of
the lines.}\label{mle_ints}
\end{figure}

Although our motivation for testing is different from the usual one
leading to the definition of the false discovery rate (FDR), the
multiple testing setup and the dependence between information matrix
tests applied successively in sliding windows should be taken into
account when we want to justify the existence of a misspecification
region. Our main interest does not lie in the discovery of regions
where the null hypothesis of well-specifiedness is rejected at a given
level of FDR [\citet{benjaminihochberg95}], but in finding those where
it cannot be rejected. Nevertheless, we tested the significance of the
departure from the null hypothesis by the procedure proposed in \citet{benjaminiyekutieli01} separately for each $u,K$ pair, and found that
the misspecification is significant for all~$K$ and for $F(u) = 0.98$
between roughly 1940--1965. The information matrix test was closest to
zero at the point $F(u) = 0.99$, $K = 4$ over the whole century, so we
chose these for smooth estimation of the extremal index. Figure~\ref{mle_ints} shows the resulting locally constant weighted $K$-gaps
estimates, compared to intervals and to iterative weighted least
squares estimates based on 41-year long sliding windows. The confidence
intervals for the $K$-gaps estimator are based on asymptotic normality.
Nonparametric bootstrap intervals were also calculated, but showed only
slight differences mostly in the middle of the century, where the
bootstrap interval was slightly wider. The value of $\theta$ dips in
the 1950s and in recent years, but overall any evidence for changes in
the clustering of summer minimum daily temperatures seems to be weak.

The information matrix test suggests the existence of fluctuations
in the time point process of extreme anomalies. Using the ten-year
moving window to de-trend the series, and a 41-year window for the
information matrix tests and the estimation of the extremal index, only
the combination of a high threshold and a relatively high run parameter
seem to yield a well-specified model. The period where the models are
misspecified roughly coincides with a local peak in the 10-year moving
median of the data set. Changes in the median temperatures may be
accompanied by changes in clustering characteristics, but perhaps using
a 10-year moving window for de-trending is insufficient to remove mean
fluctuations, so the anomalies display traces of residual
nonstationarity that then appear in the 41-year moving windows. Our
investigation thus emphasizes the importance of an appropriate
treatment of long-term trends. Many studies of climate extremes use
varying thresholds based on local quantile estimation or on an
assumption of a trend of simple parametric form, the first of which
corresponds to our ad hoc selection of window length for de-trending;
see, for example, \citet{kharinzwiers05}, \citet{nogaj06} or \citet{brownetal08}. Our results indicate that model quality is highly
sensitive to such choices, so it is necessary to check whether the
models are well specified, in order to avoid biased estimates with
underestimated variances. Climatological studies commonly directly
compare periods of a few decades, which are assumed stationary, but
this too should be performed with care. In our study, the time-scale of
the fluctuations found at extreme levels is shorter than a few decades
on thresholds $u < F^{-1}(0.99)$ in the period 1940--1965. As this is a
period where the global mean temperature based on observational data
has a marked local peak, this might arise at other stations also, and
other climate variables may also show instability on such time-scales.

%s4.2 ###
\subsection{Daily rainfall in Venezuela}

The rainfall data recorded daily between 1~January 1961 and 31 December
1999 at Maiquetia airport in Venezuela provide a striking example of
the difficulties of using simple extreme-value methods for risk
estimation. Prior to December 1999, the annual maxima of the data set
were fitted with a Gumbel model, with no diagnostics indicating a bad
fit. However, after an unusually wet fortnight in December 1999,
extensive destruction and around 30,000 deaths [\citet{larsenetal01}]
were caused by three consecutive daily precipitation totals of 120,
410.4 and 290 mm, the largest of which was almost three times greater
than the maximum of the preceding 40 years. The return period for the
peak value of 410.4 mm under simple models can be expressed in
thousands or even millions of years. Why do classical extreme-value
methods fail so catastrophically, and could more sophisticated methods
have given a different return period estimate for such an event?

\begin{figure}

\includegraphics{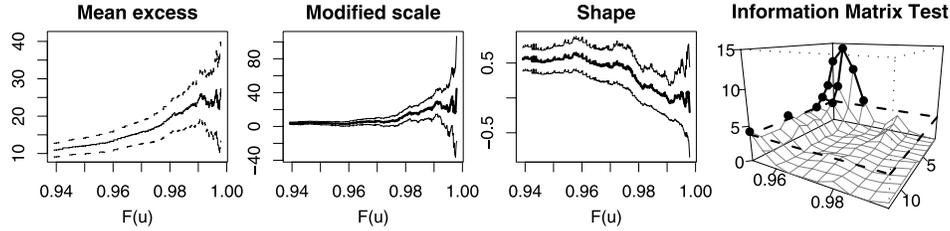}

\caption{Classical threshold
selection plots for the Venezuelan daily rainfall data for the months
December--April, between January 1961 and April 1999. The three panels
on the left show the mean excess plot and the modified scale and the
shape parameters of GPD fits as a function of threshold. The rightmost
panel is the statistic $T(\hat{\theta})$ as a function of the run
parameter $K$ and the threshold $u$ on the probability scale $F(u)$.}\label{ven_clasdiag}
\end{figure}

\citet{colespericchi03} apply a Bayesian approach to the point process
representation [\citet{smith89}], which is essentially equivalent to
fitting the generalized Pareto distribution. They use a threshold
corresponding approximately to the 0.96-quantile and including all
exceedances, and argue in favor of partitioning the sequence into two
seasons, the one of interest being from mid-November to April. With
these refinements, they obtain a predictive return period of
approximately 150 years for 410.4 mm. However, the classical threshold
selection plots for these months, shown in the left three panels of
Figure~\ref{ven_clasdiag}, indicate trouble with the model: parameter
stability is compatible with the confidence intervals only for
thresholds much higher than the 0.96-quantile. Ignoring the tendency of
extremes to cluster may also have implications for the estimates and
their variance, because the independence assumption is violated.
Moreover, which estimate should we choose if different methods give
very different answers?

Following \citet{colespericchi03} and backed by meteorological
information, we took only the months December--April, initially
excluding December 1999, and calculated our diagnostics for $u \in
[F^{-1}(0.95), F^{-1}(0.995)]$ and $K = 1, \ldots, 12$. The graph of
the statistic $T(\hat{\theta})$ in the rightmost panel of Figure~\ref{ven_clasdiag} displays three clear features: a region of
misspecification at thresholds below the 0.96-quantile; a ridge along
the 0.98-quantile with relatively higher values than the surrounding
region, corresponding to the most marked instability interval in the
classical threshold selection plots; and higher test statistic values
for $K = 1, 2$, implying misspecification for combinations with $F(u) <
0.97$, $K \leq2$. The best regions appear to be either $F(u) \approx
0.97$, $K = 3$ or $F(u) \approx0.99$, $K = 3$, but there is no further
indication which is preferable, and because of the ridge, the existence
of a contiguous area of well-specifiedness is doubtful.

%f8 ###

%f9 ###
\begin{figure}

\includegraphics{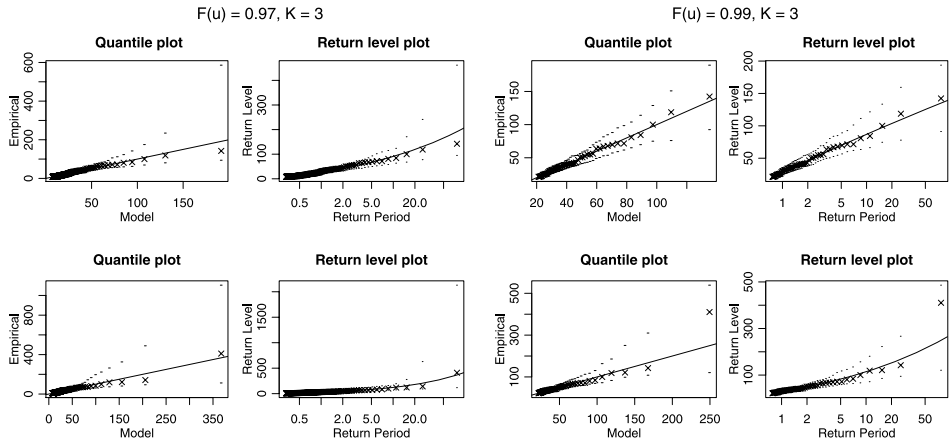}

\caption{Diagnostic and return level
plots for the generalized Pareto fits to the Venezuelan rainfall data.
The fits excluding December 1999 are shown in the top row. The left
pair of plots presents the quantile-quantile and return level plots
using the 0.97-quantile as the threshold, the right pair those using
the 0.99-quantile. The run parameter was $K = 3$ for both. The same
plots for model fits including December 1999, with the same thresholds
and run parameters, are presented in the bottom row. Each panel shows
the ordered data ($\times$) with a line representing the fitted model
and a pointwise 95\% envelope.}\label{ven_diag}
\end{figure}

The generalized Pareto model (\ref{gpd}) was fitted with $F(u) =0.97$,
$K = 3$ and $F(u) = 0.99$, $K = 3$.
The resulting parameter estimates and standard errors are very
different: $\hat{\xi} = 0.27~(0.14)$, $\hat{\sigma} = 14.8~(2.4)$ for
0.97 quantile, and $\hat{\xi} = -0.03~(0.14)$, $\hat{\sigma} =
26.6~(5.3)$ for the 0.99 quantile. Corresponding diagnostic plots are
shown in the upper row of Figure~\ref{ven_diag}, the left two plots
referring to the lower threshold, and the right two to the higher.
Neither model seems poor, but they give very different return periods
for the value 410.4 mm: approximately 600 years for the model with
threshold $F(u) = 0.97$, and an unreasonable 30 million years for the
other. The catastrophe seems to be compatible only with the lower
threshold model, despite the fact that in extreme-value statistics,
models above higher thresholds are generally considered to be closer to
the limiting distribution. Inclusion of December 1999 does not change
the misspecification tests much, but changes the estimates to $\hat{\xi
} = 0.5~(0.15)$, $\hat{\sigma} = 12.6~(2.2)$\vspace*{1.5pt} for the lower threshold
and $\hat{\xi} = 0.27~(0.15)$, $\hat{\sigma} = 25.1~(5.2)$ for the
higher one. The corresponding return periods become $65$ and $300$
years, respectively. A look at the diagnostic plots in the bottom row
of Figure~\ref{ven_diag} confirms that the former model admits the
catastrophe quite smoothly, whereas it remains an outlier in the latter.

One explanation of this apparent paradox might be that the underlying
process is a mixture. Rainfall is generated by different atmospheric
processes, such as convective storms, cold fronts or orographic winds.
If so in this case, the observed extreme process corresponds to the
extremes of a mixture distribution $\sum_{i=1}^m p_iF_i$ with the
component distribution functions $F_i$ appearing with probabilities
$p_i$, where the $F_i$ could have different extreme-value limits: with
$m=2$, for example, $F_1$ might correspond to the short-tailed case $\xi
=-1/2$ and $F_2$ to the long-tailed case $\xi=1$. Such a mixture can
show unstable behavior like that of Figure~\ref{ven_clasdiag}, which
hints at the presence of at least two components: a more frequent
light-tailed one with relatively high location and scale parameters,
dominating the levels around the 0.99-quantile, and a rarer
heavy-tailed one concentrated at lower levels and having a smaller
scale parameter, but generating extremely large observations
occasionally. A more sophisticated model for the clustering of extremes
also suggests a mixture character, but will be reported elsewhere. This
failure of simple extreme-value techniques is a warning to beware of
oversimplification, and suggests that an approach linking atmospheric
physics and statistical methods would provide better risk estimates.

%s5 ###
\section{Discussion} \label{secdisc}

Inference about the extremal behavior of a process involves assumptions
such as asymptotic independence at extreme levels and stationarity, and
also entails the selection of auxiliary quantities such as threshold
and run parameters in order to apply asymptotic models with finite
samples. Careful investigation of possible model misspecification is
therefore essential.

In this paper we have applied standard methods of detecting
misspecification to the point mass-exponential mixture model (\ref{mxtexp}) for the inter-exceedance times. These tests assist in the
selection of the threshold and the run parameter $K$ and thus help to
provide better estimates of both the extremal index and of the
generalized Pareto distribution. Failure of the model (\ref{mxtexp})
indicates failure of a more general limit, and consequently of the GPD
approximation~(\ref{gpd}). Analysis of the Venezuelan rainfall data
shows that misspecification tests can provide a valuable supplement to
classical threshold selection procedures, can lead to improved models
and better variance estimates, and may yield further insight into the
structure of the data.

We have also described a maximum likelihood estimator for the extremal
index, based on the point mass-exponential model (\ref{mxtexp}) and on
the existence of a selection procedure for $K$. The maximum likelihood
estimator is consistent and asymptotically normal under an appropriate
choice of $K$, and shows small asymptotic bias and root mean square
error compared to the best competing estimators. The joint application
of the misspecification tests and the smoothed maximum likelihood
estimator proved the good properties of the procedure as an efficient
method to detect violations of underlying assumptions such as
nonstationarity or indicate other model problems like mixture character
that cannot be disregarded using finite thresholds. It can be therefore
a useful aid to fine-tuning parameters of extreme-value models or
investigating their limitations.

One natural question is whether the assessment of misspecification for
extremal models might better be based on (\ref{gpd}). The difficulty
with this is that the $r$th moment of the score statistic for $\xi$
exists only if $r\xi>-1$, so the maximum likelihood estimators of $\xi$
and $\sigma$ are regular only if $\xi>-1/2$ [\citet{smith85}] and the
observed information has finite variance only if $\xi>-1/4$, and
information quantities for~(\ref{gpd}) have poor finite-sample
properties. The distribution~(\ref{mxtexp}), on the other hand, is
regular for $0<\theta<1$, and so has no such disadvantages.

\begin{appendix}

\section*{Appendix: Formulae for the information matrix test}\label{appendix}

Assume that $X_1,\ldots, X_n$ satisfy the necessary conditions for
Theorem~\ref{theo_Kgapconv}. For a threshold $u_n$, suppose that $N<n$
observations exceed the threshold $u_{n}$. Let the indices $\{j_i:
X_{j_i}> u_n\}$ denote the times of the exceedances, and let $c_i^{(K)}
= \overline{F}(u_n) s_i^{(K)} = \overline{F}(u_n) \max(j_{i+1} - j_{i}
- K,  0)$ denote the $i$th observed $K$-gap $s_i^{(K)}$ normalized by
the tail probability $\overline{F}(u_n)$ for $ i = 1, \ldots, N-1$ and
$K = 0, 1, \ldots.$ Then, denoting derivatives with respect to $\theta$
by a prime, it follows from the likelihood \eqref{goodloglik} that
\begin{eqnarray*}
\ell_K'  \bigl(\theta; c_i^{(K)}  \bigr) &=& - { I (c_i^{(K)} =
0 ) \over(1-\theta) } + { 2 I (c_i^{(K)} > 0 ) \over
\theta} - c_i^{(K)}, \\
\bar{J}_n(\theta) &=& (N-1)^{-1} \sum_{j = 1}^{N-1}  \biggl\{ {I
(c_j^{(K)} = 0 ) \over(1- \theta)^2} +
{4 I (c_j^{(K)} > 0 ) \over\theta^2} + c_j^{(K)} - {4
c_j^{(K)} \over\theta}  \biggr\}, \\
\bar{I}_n(\theta) &=& (N-1)^{-1} \sum_{j = 1}^{N-1}  \biggl\{ {I
(c_j^{(K)} = 0 ) \over(1- \theta)^2} +
{2 I (c_j^{(K)} > 0 ) \over\theta^2}  \biggr\},
\end{eqnarray*}
where $I(A)$ is the indicator function of the set $A$. Then we can
derive the one-observation version, the sample mean of the difference
$D$ between the variance of the score and the Fisher information and
the sample variance of the latter as
\begin{eqnarray*}
d\bigl(\theta; c_i^{(K)}\bigr) &=& {2 I (c_i^{(K)} > 0 ) \over\theta^2}
+ c_j^{(K)} - {4 c_j^{(K)} \over\theta}, \\
D_n(\theta) &=& \bar{J}_n(\theta) - \bar{I}_n(\theta),
\\
D_n'(\theta) &=& (N-1)^{-1} \sum_{j = 1}^{N-1}  \biggl\{ - { 4 I
(c_j^{(K)} > 0 ) \over\theta^3 } + { 4 c_i^{(K)} \over\theta^2 } \biggr\}, \\
V_n(\theta) &=& (N-1)^{-1} \sum_{j = 1}^{N-1}  \bigl\{ d\bigl(c_j^{(K)}\bigr) -
D_n' \bar{I}_n(\theta)^{-1} \ell_K'  \bigl(\theta; c_i^{(K)}  \bigr)
 \bigr\},
\end{eqnarray*}
and from there, the information matrix test statistic is obtained by
substituting the appropriate quantities and the estimated value of $\hat
{\theta}_n^{(K)}$ as
\[
T\bigl(\hat{\theta}^{(K)}_n\bigr) = nD_{n}\bigl(\hat{\theta}^{(K)}_n\bigr)^2 V_{n}
\bigl(\hat{\theta}^{(K)}_n\bigr)^{-1}.
\]
\end{appendix}

\section*{Acknowledgments}

We gratefully acknowledge helpful comments of Philippe Naveau, Jonathan
Tawn, two referees, an associate editor and the editor.

\printaddresses

\end{document}